\documentclass[preprint,12pt]{elsarticle}

\usepackage{graphics}
\usepackage{amssymb}
\journal{Journal of Crystal Growth}

\begin{document}

\begin{frontmatter}



\title{Validity of commonly used formula of nucleation work for bubble nucleation}


\author{Atsushi Mori\footnote{Corresponding author: E-mail atsushimori@tokushima-u.ac.jp, Tel +81-88-656-9417, Fax +81-88-656-9435}}

\address{Institute of Technology and Science, The University of Tokushima,
2-1 Minamijosanjima, Tokushima 770-8506, Japan}

\begin{abstract}
% Text of abstract
Nishioka and Kusaka [J.~Chem.~Phys. \textbf{96} (1992) 5370 (1992)] showed that the commonly used formula, $W = n(\mu^\alpha-\mu^\beta) + \gamma A$, for work of formation of critical nucleus is derived by integrating the isothermal Gibbs-Duhem relation for the incompressible nucleating phase, such as an incompressible liquid phase nucleation in a vapor phase.
In their paper as wel as in a subsequent paper [Li, Nishioka, and Holcomb, J.~Cryst.~Growth \textbf{171} (1997) 259] it was stated that the commonly used formula was valid for an incompressible nucleating phase and no longer held for such as a bubble nucleation.
In this paper, we will amend this statement; that is, the commonly used formula is shown to hold for incompressible parent phase, such as a bubble nucleation in an incompressible parent phase.
\end{abstract}

\begin{keyword}
A1 Critical nucleus formation work;
B1 Bubble nucleation;
B1 Incompressible parent phase \\
% MSC codes here, in the form: \MSC code \sep code 
% or \MSC[2008] code \sep code (2000 is the default)
\end{keyword}

\end{frontmatter}

\newpage

% \linenumbers

\noindent
\textit{Introduction} ---
In 1992, Nishioka and Kusaka~\cite{nishioka1992} showed that the volume term in the work of formation of a critical nucleus $-V^\beta (p^\beta - p^\alpha)$ can be rewritten in
\begin{equation}
\label{eq:commonvol}
-V^\beta (p^\beta - p^\alpha) = -n \sum_i x_i^\beta \Delta\mu_i,
\end{equation}
for multicomponent systems, where $\alpha$ and $\beta$ represent the parent and nucleating phases, respectively, $V^\beta \equiv 4\pi R^3/3$ with $R$ being the radius of the surface of tension is the volume of the nucleus, $p$ denotes the pressure, $n \equiv V^\beta/v^\beta$ with $v^\beta$ being the molecular volume of the $\beta$ phase, which is assumed to be constant, is the number of molecules included in the hypothetical cluster, $x_i^\beta$ the composition of component $i$ for the $\beta$ phase (the same manner for the $\alpha$ phase), and
\begin{equation}
\label{eq:deltamubeta}
\Delta\mu_i \equiv \mu_i^\alpha(T,p^\alpha,\{x_j^\alpha\}) - \mu_i^\beta(T,p^\alpha,\{x_j^\beta\}),
\end{equation}
with $\mu$ denoting the chemical potential.
Here, $\{x_j^\alpha\}$ and $\{x_j^\beta\}$ denote $\{x_j^\beta; j=1,\cdots,c\}$ and $\{x_j^\beta; j=1,\cdots,c\}$, respectively.
$\{x_j^\beta\}$ in Eq.~(\ref{eq:deltamubeta}) is the compositions of the bulk $\beta$ phase, which are determined by $\mu_i^\alpha(T,p^\alpha,\{x_j^\alpha\}) = \mu_i^\beta(T,p^\beta,\{x_j^\beta\})$ with $p^\beta \equiv p^\alpha+2\gamma/R$ and $\gamma$ being the interfacial tension.
Through Eq.~(\ref{eq:commonvol}), we have the commonly used formula for the work of formation of the critical nucleus.
\begin{equation}
\label{eq:common}
W = -n \sum_i x_i^\beta \Delta\mu_i + \gamma A,
\end{equation}
where $A \equiv 4\pi R^2$ is the area of the surface of tension.
The point is that the  pressures as an argument of the chemical potentials in Eq.~(\ref{eq:deltamubeta}) are common to $\mu^\alpha$ and $\mu^\beta$.
Therefore, $\Delta\mu$ defined by Eq.~(\ref{eq:deltamubeta}) does not coincide to the supersaturation $\mu_\alpha - \mu_\mathit{eq}$ with $\mu_\mathit{eq}$ being the chemical potential at the $\alpha$-$\beta$ equilibrium; for latter convenience we defined $p_\mathit{eq}$ as a solution to $\mu_\alpha(T,p)=\mu_\beta(T,p)$ and with $p_\mathit{eq}$ we can introduce $\mu_\mathit{eq} \equiv \mu_\alpha(T,p_\mathit{eq}) = \mu_\alpha(T,p_\mathit{eq})$.
The key is that the $\beta$ phase is incompressible.
That is, the Gibbs-Duhem relation at constant temperature is integrated for an incompressible $\beta$ phase.
Unfortunately, Nishioka and Kusaka~\cite{nishioka1992} incorrectly stated that the commonly used formula was valid for incompressible $\beta$ phases and did not hold for, for example, bubble nucleations.
In a subsequent paper~\cite{li1997} this incorrect statement remained.
The purpose of this paper is to amend this incorrect statement. \\

\noindent
\textit{Calculations} ---
We limit ourselves to the single component system for simplicity and transparency of the argument.
We start with the relation
\begin{equation}
\label{eq:isothermalGD}
\left(\partial \mu /\partial p\right)_T = v, 
\end{equation}
which is nothing other than the Gibbs-Duhem relation for the isothermal case.
We will consider two cases; one is the case treated by Nishioka and Kusaka~\cite{nishioka1992}, i.e., the case of an incompressible $\beta$ phases, and the other is the case of an incompressible $\alpha$ phase.
For the first case, let us integrate Eq.~(\ref{eq:isothermalGD}) for the $\beta$ phase for $p$ from $p^\alpha$ to $p^\beta$.
We have
\begin{equation}
\label{eq:incompbeta}
\mu^\beta(T,p^\beta) \! - \mu^\beta(T,p^\alpha) = \! \int_{p^\alpha}^{p^\beta} \! \! \! \! \! v^\beta dp = v^\beta (p^\beta \! - p^\alpha).
\end{equation}
We can eliminating $p^\beta - p^\alpha$ in the volume term using Eq.~(\ref{eq:incompbeta}) to obtain an equation corresponding to Eq.~(\ref{eq:common}):
\begin{equation}
\label{eq:Wincopbeta}
W = (V^\beta /v^\beta) \Delta\mu + \gamma A,
\end{equation}
where
\begin{eqnarray}
\nonumber
\Delta \mu &=& \mu^\beta(T,p^\alpha) - \mu^\beta(T,p^\beta) \\
\label{eq:Deltamuincompbeta}
&=& \mu^\beta(T,p^\alpha) - \mu^\alpha(T,p^\alpha).
\end{eqnarray}
To reach to the last expression, $\mu^\alpha(T,p^\alpha) = \mu^\beta(T,p^\beta)$ has been used.

For the second case, integrating Eq.~(\ref{eq:isothermalGD}) for the $\alpha$ phase for $p$ from $p^\alpha$ to $p^\beta$ and eliminating $p^\beta - p^\alpha$ in the volume term in the same way for the first case, we have
\begin{equation}
\label{eq:Wincopalpha}
W = (V^\beta /v^\alpha) \Delta\mu + \gamma A,
\end{equation}
with
\begin{eqnarray}
\nonumber
\Delta \mu &=& \mu^\alpha(T,p^\alpha) - \mu^\alpha(T,p^\beta) \\
\label{eq:Deltamuincompalpha}
&=& \mu^\beta(T,p^\beta) - \mu^\alpha(T,p^\beta).
\end{eqnarray}
One may become aware that Eq.~(\ref{eq:isothermalGD}) can be integrated for any incompressible phases to eliminate $p^\beta - p^\alpha$.
However, one should pay attention in selection of the phase in order to relate the quantities appeared with experimentally measurable ones. \\

\noindent
\textit{Discussions} ---
In this way, we have shown that the form of Eq.~(\ref{eq:common}) is also obtained for nucleations from a incompressible $\alpha$ phase, such as bubble nucleations in an incompressible liquid.
This should not be a mere exercise of theoretician; the relationship between the present $\Delta\mu$ and the experimentally measurable quantities is worth noting.
For example, for a case that the $\alpha$ phase is rarefied gas the supersaturation, which is experimentally measurable, is expressed as
\begin{equation}
\label{eq:supersaturation}
\mu_\alpha - \mu_\mathit{eq} = k_BT \ln (p_\alpha /p_\mathit{eq}).
\end{equation}
We can refer, for example, Ref.~\cite{kashchiev} for other cases of expression of the supersaturation such as the crystal-melt case.
Many experimental  results have inaccurately been analyzed with use of right-hand side of Eq.~(\ref{eq:supersaturation}) for $\Delta\mu$ of Eq.~(\ref{eq:deltamubeta}).
The correct form for an incompressible phase nucleation in a rarefied gas is given by (as done such as in Refs.~\cite{kashchiev1996,obeidat2004}) 
\begin{eqnarray}
\nonumber
&&\mu^\alpha(T,p^\alpha) - \mu^\beta(T,p^\alpha) \\
\nonumber
&=& [\mu^\alpha(T,p^\alpha) - \mu_\mathit{eq}] - [\mu^\beta(T,p^\alpha) - \mu_\mathit{eq}] \\
&=& k_BT \ln (p_\alpha /p_\mathit{eq}) - v^\beta (p_\alpha - p_\mathit{eq}),
\end{eqnarray}
which is, in principle, evaluated experimentally despite of an over idealization (the simultaneous requirement of the ideal gas and incompressible phase may be over idealization --- we note that the second  term can be omitted for limiting cases that a condensed phase nucleates in a rarefied gas).

It is thought that the procedure for the bubble nucleations in an incompressible parent phase is the same except that the pressure as an argument of the chemical potentials is the inside pressure $p^\beta$.
The pressure inside the nucleus cannot essentially be measured directly experimentally.
However, because $p^\beta$ is equal to $p^\alpha + 2\gamma/R$, the inside pressure is calculated with use of the experimentally measurable quantities.
In doing so, curvature dependence of $\gamma$ may matter.
Another attention may be on the ideal gas property of the bubble $\beta$ phase; the bubble nucleated in a liquid may not be regarded as rarefied.
Although an additional experiment is required, this problem is circumvented by using the fugacities instead of the pressures.
We have
\begin{eqnarray}
\nonumber
&&\mu^\alpha(T,p^\beta) - \mu^\beta(T,p^\beta) \\
\nonumber
&=& [\mu^\alpha(T,p^\beta) - \mu_\mathit{eq}] - [\mu^\beta(T,p^\beta) - \mu_\mathit{eq}] \\
&=& k_BT \ln (f^\beta /f_\mathit{eq}) - v^\beta (p^\beta - p_\mathit{eq}),
\end{eqnarray}
where $f^\beta=f^\beta(T,p^\beta)$ and $f_\mathit{eq}=f^\beta(T,p_\mathit{eq})$ are the fugacities corresponding to $p^\beta = p^\alpha + 2\gamma/R$ and $p_\mathit{eq}$, respectively. \\

\noindent
\textit{Summary} ---
We have amended the statement made by Nishioka and coworkers \cite{nishioka1992,li1997}.
That is, the form of the commonly used formula, Eq.~(\ref{eq:common}), has been shown to be valid for the case of an incompressible parent phase such as a bubble nucleation in an incompressible parent phase.
Moreover, we have proposed a method to relate the quantities in the obtained formula to the experimentally measurable quantities. \\

\noindent
\textit{Acknowledgment} --- The author thanks Dr. Y. Suzuki for discussions.



%% References




\end{document}